\documentclass{ws-rv961x669}

\newcommand{\beq}{\begin{equation}}
\newcommand{\eneq}{\end{equation}}
\newcommand{\be}{\begin{equation}}
\newcommand{\ee}{\end{equation}}
\newcommand{\bea}{\begin{eqnarray}}
\newcommand{\eea}{\end{eqnarray}}

\usepackage{ws-rv-van}  
\usepackage{ws-rv-thm}     
\usepackage{subfigure}     
\makeindex

\begin{document}

\chapter[Emerging Majorana Modes in Junctions of One-Dimensional Spin Systems]{Emerging Majorana Modes in Junctions of One-Dimensional Spin Systems}\label{ra_ch1}

\author[Domenico Giuliano and Andrea Trombettoni  and Pasquale Sodano]{Domenico Giuliano$^{(1,2)}$, Andrea Trombettoni$^{(3,4)}$, and Pasquale Sodano$^{(5)}$}

\address{$^{(1)}$ Dipartimento di Fisica, Universit\`a della Calabria, Arcavacata di 
Rende I-87036, Cosenza, Italy \\
$^{(2)}$ I.N.F.N., Gruppo collegato di Cosenza, 
Arcavacata di Rende I-87036, Cosenza, Italy\\
$^{(3)}$ Department of Physics, University of Trieste, Strada Costiera 11, I-34151 Trieste, Italy\\
$^{(4)}$ CNR-IOM DEMOCRITOS Simulation Center and SISSA, 
Via Bonomea 265, I-34136 Trieste, Italy\\
$^{(5)}$ I.N.F.N., Sezione di Perugia, I-64100 Perugia, Italy}

\begin{abstract}
The non-local effects induced by Majorana fermions in field theories for condensed matter systems are deeply related to the fermion charge fractionalization discovered
  by Roman Jackiw in relativistic field theories. We show how the presence of Majorana 
  fermions may be mimicked in pertinent networks of spin chains inducing a spin analogue of the multi-channel Kondo effect. The relevance of this spin analogue of the Kondo effect for
  networks of Josephson arrays and Tonks-Girardeau gases is highlighted. 
\end{abstract}


\body

\section{Introduction}
\label{intro}




Low energy neutral fermionic excitations (Majorana modes), 
deeply related to the charged fermion zero modes discovered  by
Roman Jackiw in pioneering topological investigations of relativistic field theories
\cite{JackiwRebbi,JackiwRossi}, have been claimed to be relevant in a
variety of strongly correlated condensed matter systems,
providing new insights also for the investigation of non-Fermi liquid states.

Majorana fermions were first proposed in 1937
by Ettore Majorana \cite{Majorana2008} who considered a modification to
the relativistic Dirac equation
for conventional spin-1/2 particles (Dirac fermions)
giving purely real (as opposed to
complex) solutions. These Majorana fermions are particles coinciding 
with their own antiparticles since their creation operator
is equal to their annihilation operator. In spite of the beautiful simplicity of this idea, 
Majorana fermions are not easy to come by in nature. One could, for example, 
decompose a relativistic electron, whose wave equation does have charge 
conjugation symmetry, into its real and imaginary parts. However, 
the interaction of the electron with photons is not diagonal in this 
decomposition. The real and the imaginary components will be readily 
remixed by the electromagnetic interaction: they cannot be stationary states 
of the full Hamiltonian of Quantum Electrodynamics. In condensed matter
systems, however, it may be easier to look for emergent Majorana fermions, since usually 
electromagnetic interactions are screened. 

Our particular interest in the following is in situations where the fermion spectrum 
has midgap, or zero energy, states. Already for complex electrons, mid-gap states 
give rise to fractional quantum numbers \cite{JackiwRebbi,raja,niemi,jkk}   relevant for the studies on polyacetylene \cite{js}.
With Majorana fermions, these states lead to 
peculiar representations of the anticommutator algebra, which can violate basic 
symmetries, such as fermion parity symmetry \cite{Jackiw_M00,Jackiw_M01,Jackiw_M1,wen_0,LeeWilczek}. 

The huge interest in Majorana fermions goes beyond fundamental curiosity
since there is an enormous potential for applications to
quantum technology and to devices based on the manipulation of Majorana fermions.
For example, this could allow an electron to be splitted   
in a pair of widely separated Majorana
bound states \cite{SemenoffSodano}, which could be less sensitive
to the effect of
localized sources of
decoherence. Current research attempts to develop integrated devices suitable for detecting,
storing and manipulating Majorana fermions \cite{Alicea_2012} 

Junctions of one-dimensional (1D) wires seem to possess several distinct advantages
when it comes to fabrication and subsequent detection of Majorana zero modes. In these
1D devices, zero energy Majorana modes are confined either
at the wire edges or at a domain
wall between topological and non-topological regions of the wire and,
due to the 1D confinement,  there are very few modes to
``disturb'' their observable signatures.
From the experimental side, there have been
recent advances in fabrication and manipulation of clean quantum
wires allowing for an
unprecedented level of control and analysis of these devices
in a wide range of settings.

Low energy Majorana modes have been recently the object of many theoretical
\cite{benan,Alicea_2012,franz,Leijnse_2012} and experimental \cite{Mourik1003} investigations. Located at the edges of
1D devices, they are responsible for the emergence of stretched
nonlocal electron states
\cite{kitaev,SemenoffSodano} allowing for distance independent
tunnelling \cite{Semenoff_2007}, crossed Andreev reflection \cite{been_0},
teleportation-like coherent transfer of a fermion 
\cite{fu} and fractional Josephson effects
\cite{ji,sau2012possibility}. Their effects emerge in
a variety of platforms: quantum wires
immersed in a $p$-wave superconductor \cite{kitaev,SemenoffSodano,Semenoff_2007}, cold atomic systems \cite{tewari},
topological insulator-superconductor magnetic structures \cite{fu,fukane,shi},
semiconductor heterostructures \cite{sau_1,sau_2,oreg,dassarma},
superconducting wires \cite{tsvelik2011zero}, Josephson arrays
\cite{Hassler_2012} and spin
systems \cite{Nersesyan_2011} In addition,
they may be relevant excitations also in conventional $s$-wave
superconductors \cite{PhysRevB.81.224515}.

It is worth to stress here that, when a superconductor is coupled to a conducting wire in a SN-junction, the   Majorana mode hybridizes with the conducting electrons at the normal side of the junction. Such a feature eventually yields to a "Majorana hybridization" with the electrons in the normal wire 
\cite{simon}, a phenomenon strikingly similar with the emergence of the "electronic Kondo cloud" at the Kondo fixed point \cite{afgiu}.

Majorana low energy modes are expected to be relevant for applications to
topological quantum
computing \cite{RevModPhys.80.1083} and quantum interferometry \cite{Bose_2011,strubi} and their
manipulation is under current investigation \cite{fukane_2,ak,roma,KITAEV20032,RevModPhys.80.1083}. In junctions of quantum wires, 
Majorana edge modes induce, for various network topologies,
remarkable even-odd effects on the tunneling conductance \cite{Zazunov_2013}.
In addition, Coulomb charging effects cause
conductance oscillations and resonances connected to teleportation as
well as peculiar finite-bias peaks \cite{PhysRevB.84.165440,PhysRevLett.109.166403} and
may trigger the flow towards exotic Kondo fixed points
when the center island has a finite charging energy
\cite{PhysRevLett.110.196401}.
 
Majorana fermions in condensed matter are not fundamental particles. Rather, they
are effective degrees of freedom, emerging in the presence of a degeneracy in the
ground state which is topologically protected. One paradigmatic example is the Kitaev
chain \cite{kitaev}, a tight-binding model for the effectively spinless fermions in 1D $p$-wave superconductor.
In this model, a topologically protected
phase (i.e., a phase which cannot be changed by any local operation, and thus robust against
interaction with the environment) can be described in terms of a pair of Majorana modes localized at the ends of the chain.
Together, they form a non-local fermionic degree of freedom, which can be used to
encode a qubit. As a solid-state realization of this model, a set of nanowires with strong Rashba
coupling (InAs, InSb), laid on a conventional BCS superconductor (Al, Nb) and subject
to a suitably tuned magnetic field can develop Majorana ending modes \cite{oreg,dassarma}, for which
experimental evidence has been provided (e.g. \cite{Mourik1003}).

\begin{figure}[h!]
 \centering
 \includegraphics[width=0.6\textwidth]{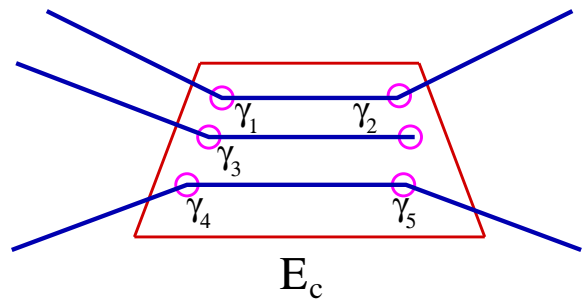}
 \caption{Schematic representation of the topological Kondo model:
   Majorana edge modes of quantum wires are hosted on an $s$-wave superconductor
   and are capacitively coupled with the ground.
   These modes are proximity-coupled with the external leads.}
\label{fig:MajoranaBox}
\end{figure}

The setup described above is used in the so-called Majorana-Coulomb box, or topological
Kondo model   (TKM)   \cite{BeriCooper2012}. It is obtained connecting a set of $M$ effectively 1D wires to a set of
nanowires supporting Majorana modes at their ends, hosted on a mesoscopic
superconducting substrate with a large charging energy, and subject to an applied
voltage potential (see Fig.\ref{fig:MajoranaBox}). 
At low temperatures, the model is integrable \cite{Altland2013},
despite not being 1D. It is possible, via conformal field theory or Bethe ansatz, to compute
thermodynamic quantities
such as free energy, specific heat and entropy contribution from the central region \cite{Buccheri2015}.

It should be noticed that conventional multi-channel Kondo models are difficult to attain \cite{bay_1}, due to the need for a
perfect symmetry between the couplings of the spin density from the channels to
the spin of the impurity.
The TKM provides a shortcut to circumvent this problem \cite{BeriCooper2012}, relating the symmetry between the channels to the
topological degeneracy characterizing a 1D quantum wire
into a topological phase, and associated with the emergence of localized
Majorana modes at the edges of the wire.

Neglecting the overlap between Majorana modes lying at the edges of the wire,
the authors of \cite{BeriCooper2012} showed that the ground state of the wire exhibits a topological
degeneracy and that, on coupling $M$ topological quantum wires to a central
superconducting island with a finite charging energy, for $M \geq 3$, one may realize over
screened multi-channel Kondo physics, with the spin of the isolated impurity given by
a pertinent combination of the real fermionic operators of the Majorana
modes localized at the inner edges. The emerging Kondo Hamiltonian is topological as
the central spin is a nonlocal combination of emerging Majorana modes. This allows for
a robust realization of multi-channel Kondo physics where the effects of the potentially
dangerous anisotropies in the couplings between the spin impurity and the various
channels are suppressed due to topology, being irrelevant in the renormalization group flow
towards the Kondo regime \cite{BeriCooper2012,tsve_1}.

Finally, one should observe that the competition between superconductivity and Kondo physics leads to unusual Josephson current-phase relations and to a remarkable transfer of fractionalized charges \cite{zaz_1}.
 Also interesting is the study of the concurrence of entanglement between two quantum dots in contact with Majorana bound states on a floating superconducting island. One finds that the distance between the Majorana states, the charging energy of the island and the average island charge are decisive parameters for the efficiency of the entanglement generation. This leads to the possibility of long range entanglement with distance independent concurrence over wide parameter regions \cite{plug}.

A different platform for the multi-channel Kondo model
has been recently proposed in \cite{crampettoni} for a $Y$-junction of three 1D $XX$ models and
by Tsvelik in a $Y$-junction of three 1D quantum Ising models, in both cases with the 1D chains 
joined at their inner edges \cite{tsve_1}.
This proposal, when the relevant parameters are pertinently tuned, is particularly attractive since
the behaviour of  the uncoupled spin models is known and it provides reliable and effective descriptions of strongly correlated
phenomena in condensed matter systems. As a result one may envision to probe the
multi-channel Kondo effect in a variety of controllable, and yet ``topologically robust'',
experimental settings such as the ones provided by degenerate Bose
gases confined in an optical lattice \cite{Buccheri15_bis,PhysRevA.96.033603}
or quantum Josephson junction networks  
(Refs.\cite{ChamonPhysRevLett.91.206403,chamon,giuso_08} and Refs. therein).
It should be noticed that spin realizations of multi-impurity \cite{Bay12} and multi-channel \cite{alkur} Kondo models revealed very useful 
not only for quantifying the entanglement \cite{Bay10}, but also for characterizing quantum phase transitions \cite{nature_psetal}.

Remarkably, the TKM behavior is that of a non-Fermi liquid. Configurations based
on plaquette schemes are probably the simplest way of realizing a topologically
protected memory \cite{PhysRevLett.116.050501}, and, within a surface code, all operations required for quantum
computation. These configuration can be realized using as basic unit the Majorana-Coulomb box \cite{tsve_3}.
Remarkably, the TKM can be realized by a junction of spin
chains \cite{Buccheri15_bis} and it could be obtained in laboratory with the use of holographic optical
traps for cold atomic gases, which paves the way of a whole new family of
holographic devices \cite{Buccheri15_bis}.

The plan of the paper is the following. Using a pertinently generalized Jordan-Wigner transformation, 
Section \ref{sec:qsc} shows how a $Y$-junction of $XY$ spin chains may be transformed in a fermionic model with a 
non-trivial boundary term containing 
Klein factors (KF), i.e.  
real fermion modes, introduced to 
ensure the appropriate commutation relations on the Y-graph. Section \ref{sec:the} is meant to clarify how the 
emergence of real fermion KF opens the possibility to engineer a spin-chain version of the TKM.
In addition, one sees that a $Y$-junction of quantum Ising chains leads to a two-channel 
version of the topological Kondo effect, while a $Y$-junction of $XX$ spin chains leads to a four channel 
Kondo model.  Section \ref{jjn} shows how one can map networks of quantum spin chains -- in a pertinent range 
of parameters -- in networks of Josephson junction arrays. In Section \ref{sec:hol} one sees how a Y-junction of $XX$ 
spin chains may be used to describe cold atomic systems in holographic traps. Section \ref{sec:conc} summarizes our findings.

To help following the various abbreviations, we list in Table 
\ref{abbr} the meaning of  the ones we use most commonly throughout the paper.

\vspace{0.2cm}

 \begin{table}[!ht]
\tbl{Glossary of most commonly used abbreviations.}
{\begin{tabular}{| c | c |}
\hline 
1D & One-dimensional\\
\hline
TKM & Topological Kondo model    \\
 \hline
KF & Klein factors  \\
 \hline
QIC &  Quantum Ising chain   \\
 \hline
JW & Jordan-Wigner    \\
 \hline
YSC &  Y-junction of spin chains   \\
 \hline
JJC & Josephson junction chains  \\
 \hline
TG & Tonks-Girardeau  \\
 \hline 
\end{tabular}
 \label{abbr}
}\end{table}
\noindent

\section{Emerging Majorana modes at junctions of quantum spin chains}
\label{sec:qsc}

In this Section we address networks of spin chains with nearest-neighbour
magnetic exchange terms 
in the $x$- and $y$-directions in spin space (not necessarely equal to each other); in 
addition, we alledge for a nonzero uniform magnetic field applied in the $z$-direction. 
Specifically, each arm of the network is described by an $XY$ spin-1/2  Hamiltonian given by 

\beq
H_{XY}= - J \sum_{ j = 1}^{\ell - 1 } (S_j^x S_{j+1}^x + \gamma S_j^y S_{j+1}^y ) 
- H \sum_{ j =1}^\ell S_j^z 
\:\:\:\: . 
\label{xy.1}
\eneq
\noindent
In Eq.(\ref{xy.1}) $\vec{S}_j$ is a quantum spin-1/2 operator acting at site-$j$ of 
an $\ell$-site chain (with $\ell$ kept finite and eventually
sent to infinite at the end of the calculations), obeying the algebra 

\beq
[ S_j^a , S_{j'}^b ] = i \delta_{j , j'}\: \epsilon^{a b c } S_j^c 
\:\:\:\: .
\label{xy.2}
\eneq
\noindent
$J$ is the magnetic exchange strength between spins, $H$ is the applied, uniform magnetic field in 
the $z$-direction and $\gamma$ is the anisotropy parameter of the spin exchange interaction.
In the following $\gamma$ is chosen so that $0 \leq \gamma \leq 1$. By tuning $\gamma$ between the 
extreme values $\gamma = 0$ and $\gamma = 1$, the $H_{XY}$ in (\ref{xy.1}) continuosly interpolates between 
the quantum Ising chain   (QIC)  and the $XX$ chain  in transverse magnetic field.

To  describe a junction of several $XY$ chains, we introduce the boundary 
Hamiltonian $H_\Delta$, given by
 
\beq
H_\Delta= - J_\Delta \sum_{\alpha \neq \beta}(S^x_{1,\alpha} S_{1,\beta}^x + \gamma S_{1,\alpha}^y S_{1,\beta}^y ) 
\:\:\:\: . 
\label{xy.delta}
\eneq
\noindent
In Eq.(\ref{xy.delta}) the $j$-th spin of the $\alpha$-th chain (with $\alpha=1,\cdots,M$) is denoted
by $\vec{S}_{j,\alpha}=(S^x_{j,\alpha},S^y_{j,\alpha},S^z_{j,\alpha})$. In $H_\Delta$,
each spin in position $j=1$ of a chain $\alpha$ is coupled with all the others spins in the position
$j=1$ of all other chains $\beta \neq \alpha$. The advantage of the form (\ref{xy.delta}) is that one can safely take the continuous limit
in each chain, and the wires are coupled by a tunneling term. 

The standard approach to the single chain described by the Hamiltonian (\ref{xy.1}) consists in 
mapping $X_{XY}$ onto a quadratic, spinless fermion Hamiltonian via the Jordan-Wigner   (JW)  
transformations \cite{jordanwigner,lieb61}. The latter allows for rewriting the quantum spin operators 
$S_j^z$ and  $S^\pm_j$, where  $S^\pm_j=  S_j^x \pm i S^y_j $,  
in terms of spinless lattice operators $\{ a_ j, a_j^\dagger \}$ as 

\begin{eqnarray}
 S_{ j  }^+ &=&  a_{ j  }^\dagger 
 e^{ i \pi \sum_{ r = 1}^{j-1} a_{ r  }^\dagger a_{ r  } }  \nonumber \\
 S_{ j  }^- &=&  a_{ j  } e^{ i \pi \sum_{ r = 1}^{j-1} a_{ r  }^\dagger 
 a_{ r  } }   \nonumber \\
 S_j^z &=& a_j^\dagger a_j - \frac{1}{2}
\;\;\;\; .
\label{xy.3}
\end{eqnarray}
\noindent
In terms of the JW-fermions, one  obtains 

\beq
H_{XY} = - \frac{ J ( 1 + \gamma  ) }{2} \sum_{ j = 1}^{\ell - 1 }
\{ a_j^\dagger a_{j + 1} + a_{j + 1}^\dagger a_j \} 
+  \frac{ J ( 1 - \gamma )}{2} \sum_{j = 1}^{\ell - 1}
\{ a_j a_{j + 1 } + a_{ j +1}^\dagger a_j^\dagger \}
+ H \sum_{ j = 1}^\ell a_j^\dagger a_j 
\;\;\;\; .
\label{xy.4}
\eneq
\noindent
Once written in terms of JW-fermions, from Eq.(\ref{xy.4}) one has that,
for $\gamma = 1$,  
$H_{XY}$ reduces to the Hamiltonian for free fermions on a 1D lattice with
a chemical potential $-H$ in the $XX$-limit, and to the 
Hamiltonian for a 1D Kitaev model for a $p$-wave 
superconductor when $\gamma=0$ \cite{kitaev}. 

Remarkably, while the above JW transformations (\ref{xy.3}) are 
sufficient to rewrite the generic XY-Hamiltonian for a single chain 
in fermionic coordinates, a fundamental problem arises when several 
chains are connected to each other to form a junction of spin chains.
Of course, one should have $M \geq 3$, since a junction of two chains can always be mapped onto one chain.
Thus, for the sake of simplicity, in the following  we
shall take $M=3$. The network with $M=3$ is called a $Y$-junction of 
spin chains.  

In general, models of $Y$-junctions have been studied at the crossing
of three, or more, Luttinger liquids 
\cite{Kom,lal02,chamon03,Oshikawa_2006,giuliano_08,giuso_08,agarwal14,mardanya15,giuliano15,berin1}, in 
Bose gases in star geometries \cite{Bur01,Bru04,Tok}, and in
superconducting Josephson junctions \cite{giuliano_08,giuso_08,Cirillo}. 
When introducing more than one chain, the Hamiltonian of the uncoupled chains
(henceforth referred to as the ''leads''), $H^{(0)}$,
will be given by the sum of many Hamiltonians like 
the one in Eq.(\ref{xy.4}). In the specific case 
of the Y-junction of spin chains   (YSC) , one has
\beq
H=H^{(0)}+H_\Delta
\:\:\:\: , 
\label{xy.5_tot}
\eneq
\noindent
where $H_0$ is the bulk Hamiltonian of the uncoupled chains and $H_\Delta$ is the junction Hamiltonian. The bulk Hamiltonian is given by
\beq
H^{(0)}= \sum_{\alpha = 1,2,3} 
\left\{- J \sum_{ j = 1}^{\ell - 1 } (S_{j , \alpha}^x S_{j+1 , \alpha }^x + \gamma S_{j , \alpha}^y S_{j+1 , \alpha}^y ) 
- H \sum_{ j =1}^\ell S_{j , \alpha}^z  \right\}
\equiv \sum_{\alpha = 1,2,3} H_{XY ; \alpha }
\:\:\:\: , 
\label{xy.5}
\eneq
\noindent
with $\alpha$ being the chain index, while the junction Hamiltonian $H_\Delta$ is given by (\ref{xy.delta}). 
Furthermore, in order to obtain the  Kondo effect in spin chains, 
one should  assume $| J_\Delta | / |J| \leq 1$. 

The YSC-Hamiltonian $H \equiv H^{(0)} + H_\Delta$ provides the natural generalization 
of the Hamiltonian for a Y-junction of QIC's introduced in Ref.\cite{tsve_1}  and 
later on generalized in Refs.\cite{tsve_2,tsve_3}, to which it reduces 
for $\gamma = 0$, as well as of the  
Hamiltonian for a Y-junction  of three quantum $XX$-spin chains
discussed in Ref.\cite{crampettoni},
to which it reduces for $\gamma = 1$.

In view of the successfull application of the JW-transformation to solving a single 
$XY$-chain, one may attempt to look for an appropriate generalization to the YSC 
of the mapping onto a spinless fermionic model, 
but a simple-minded generalized JW-representation of the spin coordinates such as 

\begin{eqnarray}
 S_{ j  , \alpha  }^+ &=&  a_{ j  , \alpha  }^\dagger 
 e^{ i \pi \sum_{ r = 1}^{j-1} a_{ r  , \alpha  }^\dagger a_{ r  , \alpha  } }  \nonumber \\
 S_{ j , \alpha   }^- &=&  a_{ j  , \alpha  } e^{ i \pi \sum_{ r = 1}^{j-1} a_{ r  , \alpha  }^\dagger 
 a_{ r  , \alpha  } }   \nonumber \\
 S_{j , \alpha}^z &=& a_{j , \alpha}^\dagger a_{j , \alpha} - \frac{1}{2}
\;\;\;\; ,
\label{xy.7}
\end{eqnarray}
\noindent
would eventually lead to operators that properly commute with each other, if they belong to 
the same chain, but such that $ \{ S_{j , \alpha}^\pm , S_{j' , \alpha'}^\pm \} = 0$, if 
$\alpha  \neq \alpha'$, instead than $ [  S_{j , \alpha}^\pm , S_{j' , \alpha'}^\pm ] = 0$.

The solution to this problem was first put forward in 
Ref.\cite{crampettoni}, in analogy to what is done when bosonizing spinful 
fermionic operators in interacting one-dimensional electronic systems \cite{vondelft}.
Basically, one introduces the so-called KF, that is, real 
fermion modes, one per each chain. In the specific case of the 
YSC, one has to introduce three KF,  
$\eta^1 , \eta^2 , \eta^3$, obeying the appropriate anticommutation relations with 
each other and anticommuting with all the other JW-fermion operators used to 
represent the spin operators in the leads. Specifically, one sets 

\begin{eqnarray}
&& \{ \eta^\alpha , \eta^{\alpha'} \} = 2 \delta_{ \alpha , \alpha'} \nonumber \\
&& \{ \eta^\alpha , a_{j , \alpha'} \} = \{\eta^\alpha , a_{j , \alpha'}^\dagger \} = 0 
\:\:\:\: , 
\label{xy.8}
\end{eqnarray}
\noindent
with $\alpha , \alpha' = 1,2,3$.

Using the KF, one  modifies Eqs.(\ref{xy.7}) so  as to  take into account 
the commutation relations between the spin operators; that is, one sets \cite{crampettoni,tsve_1,gstt}

\begin{eqnarray}
 S_{ j  , \alpha  }^+ &=&  i  a_{ j  , \alpha  }^\dagger 
 e^{ i \pi \sum_{ r = 1}^{j-1} a_{ r  , \alpha  }^\dagger a_{ r  , \alpha  } }  \eta^\alpha  \nonumber \\
 S_{ j , \alpha   }^- &=& i  a_{ j  , \alpha  } e^{ i \pi \sum_{ r = 1}^{j-1} a_{ r  , \alpha  }^\dagger 
 a_{ r  , \alpha  } }   \eta^\alpha  \nonumber \\
 S_{j , \alpha}^z &=& a_{j , \alpha}^\dagger a_{j , \alpha} - \frac{1}{2}
\;\;\;\; .
\label{xy.9}
\end{eqnarray}
\noindent
Apparently, the KF are introduced as a mathematical tool to ensure the correct commutation 
relations between operators acting on sites belonging to different chains. Equivalently, one may 
just state that their introduction corresponds to an ''artificial'' enlargement of the Hilbert space,
which is yet necessary, in order to recover the correct operator algebra
\cite{crampettoni}. 

When joining together the leads into 
the YSC, the KF conspire to actually interact with the dynamical degrees of freedom of the leads,
thus being legitimately promoted to ''physical'' degrees of freedom of the system, whose presence, or 
absence, can potentially strongly affect the behavior of the junction. Thus, in 
this respect, they can definitely be regarded as ''emerging real-fermion degrees of freedom'', that is, 
as emerging Majorana modes, by no means different from the ''actual'' Majorana modes emerging in 
e.g. a 1D topological superconductor in its topological phase \cite{kitaev}. 
Notice that a number of exotic phases has been predicted to emerge in pertinently designed junctions
of 1D leads, due to the effect of coupling the ''bulk'' degrees of freedom of the system 
to the KF \cite{Oshikawa_2006}, to the so-called Klein-Majorana hybridization between the 
two kinds of real-fermion degrees of freedom entering the junction Hamiltonian \cite{beri}, or 
to both phenomena taking place at the same time \cite{ga_npb}.  

Once the generalized JW transformations, Eqs.(\ref{xy.9}), are inserted into 
$H^{(0)}$ in Eq.(\ref{xy.5}), one obtains the fermionic version of 
the disconnected lead Hamiltonian as 

\begin{eqnarray}
H^{(0)} &=& \sum_{ \alpha  = 1,2,3}\: \Biggl\{ 
 - \frac{ J ( 1 + \gamma  ) }{2}\sum_{ j = 1}^{\ell - 1 }
\{a_{j ,\alpha }^\dagger a_{j + 1 , \alpha } + a_{j + 1, \alpha }^\dagger a_{j , \alpha } \}  \nonumber \\
&+&  \frac{ J ( 1 - \gamma )}{2}\sum_{j = 1}^{\ell - 1}
\{ a_{j , \alpha } a_{j + 1 , \alpha  }+ a_{ j +1 , \alpha }^\dagger a_{j , \alpha}^\dagger \}
+ H \sum_{ j = 1}^\ell a_{j , \alpha}^\dagger a_{j , \alpha }\Biggr\}
\:\:\:\: .
\label{xy.10}
\end{eqnarray}
\noindent
We see that the KF do not appear in the ''bulk'' lead Hamiltonian in Eq.(\ref{xy.10}).
At variance, they do appear in the junction Hamiltonian $H_\Delta$ in Eq.(\ref{xy.delta})
which, once rewritten in fermionic coordinates, takes the form 

\beq
H_\Delta = 2 J_\Delta \, \left( \vec{\Sigma}_1 + \gamma  \vec{\Upsilon}_1 \right) 
\cdot \vec{\cal R}
\:\:\:\: .
\label{xy.11}
\eneq
\noindent
In particular, the KF determine the ''topological'' spin-$1/2$ operator 
$\vec{\cal R}$ \cite{crampettoni,tsve_1}, which is given by 

\beq
\vec{\cal R} =  
- \frac{i}{2} \: \left( \begin{array}{c}
                       \sigma^2 \sigma^3 \\ 
\sigma^3 \sigma^1 \\ \sigma^1 \sigma^2                        
                      \end{array} \right)
\:\:\:\: . 
\label{xy.12}
\eneq
\noindent
It is, indeed, a straightforward check to verify that the components of $\vec{\cal R}$ satisfy the 
appropriate $su (2)$-commutation relation for a spin-1/2 angular-momentum operator. 
Following the notation of Ref.\cite{gstt}, in Eq.(\ref{xy.12}) we have also 
introduced the lattice spin- and isospin-density operators, $\vec{\Sigma}_j , \vec{\Upsilon}_j$, defined as
\beq
\vec{\Sigma}_j = 
- \frac{i}{2} \: \left( \begin{array}{c}
( a_{j,2} + a_{j,2}^\dagger)    ( a_{j,3} + a_{j,3}^\dagger) \\ 
( a_{j,3} + a_{j,3}^\dagger)    ( a_{j,1} + a_{j,1}^\dagger) \\ 
( a_{j,1} + a_{j,1}^\dagger)    ( a_{j,2} + a_{j,2}^\dagger)
       \end{array} \right) \;\;\; , \;\;  
\vec{\Upsilon}_j =  
 \frac{i}{2} \: \left( \begin{array}{c}
 ( a_{j,2} - a_{j,2}^\dagger)      ( a_{j,3} - a_{j,3}^\dagger)   \\ 
  ( a_{j,3} - a_{j,3}^\dagger)   ( a_{j,1} -  a_{j,1}^\dagger)   \\ 
  ( a_{j,1} -  a_{j,1}^\dagger)       ( a_{j,2} - a_{j,2}^\dagger)  
       \end{array} \right) 
\;\;\;\; . 
\label{xy.13}
\eneq
\noindent
Eqs.(\ref{xy.10},\ref{xy.11}) are the key result of the generalized JW-transformations applied to 
a YSC. They evidence the emergence of actual degrees of freedom at the junction
(the topological spin $\vec{\cal R}$) which, despite being determined by operators originally 
introduced as a mathematical mean to assure the consistency of the JW-transformations, 
eventually become true ''physical'' degrees of freedom

\section{Kondo effect in spin systems and the topological Kondo model}
\label{sec:the}

The most striking physical consequences of the emergence of real-fermion KF at a junction of 
bosonic quantum spin chains is the possibility of using them to engineer spin-chain version of the 
so-called  Topological Kondo effect.

Historically, the Kondo effect was discovered as a low-temperature upturn in the resistivity of 
conducting metals doped with magnetic impurities  \cite{kon,Hew93,Kou01}. It
is determined by the relevance of nonperturbative spin-flip processes due to 
the magnetic dipole interaction between  the spin of a magnetic impurity and of the itinerant conduction electrons 
in the metal, which, as the temperature goes down, determines  the formation 
of a strongly correlated Kondo state between the impurity and the conduction electrons (the
''Kondo cloud'') \cite{kon,Hew93,Noz1974}. 

In the last decades, the development in quantum device 
fabrication techniques has triggered a novel interest in the Kondo effect, as it has been possible to realize it   
in quantum dots with metallic \cite{Ali96,Kou98,gg98_1,gg98_2}, 
as well as superconducting leads \cite{avish,choi,gbab}.

Of particular relevance  is the 
proposal of realizing the Kondo impurity spin in terms of  Majorana fermion modes
emerging at the endpoints of one-dimensional quantum wires connected to topological superconductors. 
Such a peculiar realization of Kondo effect is characterized by the ''topological'' nature of 
the impurity spin, each component of which is determined by Majorana modes belonging to 
different quantum wires (leads). Thus, we see that in such a topological version of 
Kondo effect (the TKM), the ''emerging'' magnetic impurity is nonlocal in the lead 
index. Therefore,  disconnecting even a single lead from the junction fully destroys the 
effect, which is the main signature of the topological nature of the phenomenon \cite{beri12,Altland2013,Buccheri2015,topo_4}.

In Refs.\cite{BeriCooper2012,AltlandEgger}, it was shown that networks of quantum wires supporting edge Majorana
modes \cite{ZazunovEvenOdd} provide a possible experimentally attainable realization of the TKM.
Subsequently, the spin dynamics \cite{Altland2013}, as well as the exact
solution for finite number of electrons \cite{Altland2014}, were investigated.

Tipically, a TKM is realized at a junction of quantum wires connected to a 
superconducting island with a finite charging energy $E_c$. In such a system, 
Coulomb interactions play an essential role for the host superconducting region,
since they determine $E_c$ \cite{beri12,Altland2013}. Focusing onto the temperature regime $T\ll E_c$,
one obtains that in all the allowed physical processes at the junction the number
of electrons on the island is conserved.
Under these conditions, the effective Hamiltonian describing the
TKM at low energy scales can be written in the form \cite{beri12,AltlandEgger,Beri2013,Zazunov14}

\beq
\label{eq:TopologicalKondoHamiltonian}
 H=-i   v_{F} \sum_{\alpha=1}^{M}\intop dx \psi_{\alpha}^{\dagger}(x)\partial_{x}\psi_{\alpha}(x)
 + \sum_{\alpha\ne\beta}\lambda_{\alpha\beta}\gamma_{\alpha}\gamma_{\beta}\psi_{\alpha}^{\dagger}(0)\psi_{\beta}(0)
 +i\sum_{\alpha\ne\beta}h_{\alpha\beta}\gamma_{\alpha}\gamma_{\beta}
 \;.
\eneq
\noindent
In Eq.(\ref{eq:TopologicalKondoHamiltonian}), the $\psi_{\alpha}(x)$'s are the
Fermi fields describing electrons in the wires $\alpha=1,\ldots,M$.
$ \gamma_{\alpha}=\gamma_{\alpha}^\dagger$ are Majorana fields constrained
in a box connected with the wires and satisfying the Poisson algebra
 \begin{equation}
  \left\lbrace\gamma_{\alpha},\gamma_{\beta}\right\rbrace = 2 \delta_{\alpha,\beta}
  \;.
 \end{equation}
 \noindent 
The symmetric matrix $\lambda_{\alpha,\beta}>0$ is the analog of the coupling with the magnetic impurity in the usual Kondo problem.
The couplings $h_{\alpha,\beta}=h_{\beta,\alpha}$ represent a direct coupling between only a pair of Majorana fermions.
These can be made exponentially small when the Majorana zero modes are 
far enough from each other in real space. For this reason, we shall  neglect them henceforth.
A related, yet different model, with real spinless fermions in the bulk, has been analyzed in \cite{TsvelikIsing}.

\subsection{Topological Kondo model from a Y-junction of one-dimensional spin systems}
 
To prove the mapping between a YSC of three critical chains and a topological Kondo 
Hamiltonian with $M=3$, one starts from  Ref.\cite{crampettoni}  where,  
by introducing the KF approach to the generalized JW transformation for a junction of 
quantum spin chains, it was shown that  
a Y-junction of quantum $XX$-spin chains in their gapless phase (corresponding to taking the
limit $\gamma = 1$ in $H^{(0)} + H_\Delta$) hosts a remarkable realization of the four-channel, spin-1/2
Kondo effect.

 In the low-energy, long-wavelength limit 
for the lead excitations, the Y-junction of quantum $XX$-spin chains exactly maps onto the Hamiltonian in 
Eq.(\ref{eq:TopologicalKondoHamiltonian}) for $M=3$. 
In order to see this, one first of all notices  that, for $\gamma = 1$, one obtains 

\begin{eqnarray}
&& H_{\gamma = 1} = H^{(0)}_{\gamma = 1} + H_{\Delta , \gamma = 1} = \\
&& \sum_{\alpha = 1,2,3} 
\left\{ - J \sum_{j = 1}^{\ell - 1} [ a_{j , \alpha}^\dagger a_{j + 1 , \alpha } + a_{ j + 1 , \alpha}^\dagger a_{ j , \alpha } ] 
-  H \sum_{ j = 1}^\ell a_{j , \alpha}^\dagger a_{j , \alpha } \right\}
+ 2 J_\Delta [ \vec{\Sigma}_1 + \vec{\Upsilon}_1 ] \cdot \vec{\cal R}  \nonumber 
\:\:\:\: . 
\label{4ck.1}
\end{eqnarray}
\noindent
In the ''critical'' region, $ 2 J > | H |$, the lattice fermions in Eq.(\ref{4ck.1}) show a gapless spectrum (in the 
$\ell \to \infty$-limit), thus behaving as gapless free fermions on the leads of the junction.  This allows  to expand the 
lattice fermion operators by only retaining low-energy modes around the Fermi momenta of the single-fermion 
spectrum in each lead, $\pm k_F = \pm {\rm arccos} \left( \frac{H}{2 J } \right)$. Specifically, one sets 

\beq
a_{j , \alpha } \approx \sqrt{a} \: \{ e^{ i k_F( j - 1) } \psi_{R , \alpha } (x_j ) + e^{ - i k_F ( j - 1 )  } \psi_{L , \alpha} ( x_j ) \}
\:\:\:\: , 
\label{x.1}
\eneq
\noindent
with $x_j = j a $ and $a$ being the lattice step (which we will set to $1$ henceforth, unless specifically required to do otherwise for the sake of 
clarity). In terms of the continuum fermion operators at the right-hand side of Eq.(\ref{x.1}), 
one  obtains 

\begin{eqnarray}
H_{\gamma = 1}  &\approx&   - i \hat{v}_F \: \sum_{\alpha = 1}^3 \: \int_0^\ell \: dx \: \{ \psi_{R , \alpha}^\dagger (  x  ) \partial_x 
\psi_{R , \alpha } ( x ) - \psi_{L , \alpha}^\dagger ( x ) \partial_x \psi_{L , \alpha } ( x ) \}  \nonumber \\
&+& \sum_{\alpha \neq \beta}  J_\Delta \sigma^\alpha \sigma^\beta  [ \psi_{R , \alpha}^\dagger ( 0 ) + 
\psi_{L , \alpha}^\dagger ( 0 ) ] [ \psi_{R , \beta } ( 0 ) + \psi_{L , \beta} ( 0 ) ] 
\:\:\:\: , 
\label{mapping.2}
\end{eqnarray}
\noindent
with $\hat{v}_F = 2 J \cos ( k_F )$. 
To recover the Hamiltonian in Eq.(\ref{eq:TopologicalKondoHamiltonian}), 
one should  ''unfold'' the chiral fields by defining 3 right-handed field $\psi_\alpha ( x )$ such that 

\beq
\psi_\alpha ( x ) = \Biggl\{ \begin{array}{l} \psi_{R , \alpha } ( x ) \;\;\; , \;\; (x>0 ) \\
\psi_{ L , \alpha } ( - x ) \;\;\; , \;\; (x < 0 ) \end{array} 
\;\;\;\; . 
\label{mm.1}
\eneq
\noindent
Once $H_{\gamma = 1}$ is rewritten in terms of the unfolded fields in Eq.(\ref{mm.1}), by comparison with 
Eq.(\ref{eq:TopologicalKondoHamiltonian}), one readily sees that $H_{\gamma = 1}$ corresponds to an $M=3$ TKM, provided 
one identifies the Majorana modes $\{ \gamma_\alpha  \}$ with the 
KF $ \{ \sigma^\alpha \}$ and $\mu$ with $H$;  all the $\lambda_{\alpha \beta}$ should be   independent 
of $\alpha$ and $\beta$ and all equal to $J_\Delta$. 

To pertinently complement our discussion about the realization of the TKM with  a YSC, it is worth recalling that 
a different spin-chain realization of the topological Kondo effect may  be realized at a Y-junction of 
three quantum Ising chains. Indeed, using the approach highlited above, in his paper onto a YSC of three critical 
quantum Ising chains \cite{tsve_1}, Alexei Tsvelik provided a remarkable realization of 
a two-channel version of the topological Kondo effect  \cite{beri12}. Upon setting
$\gamma = 0$ in Eq.(\ref{xy.10}) and, at the same time, assuming that $H$ is 
tuned at the quantum-critical point $H=J$, Eq.(\ref{xy.10}) becomes 

\beq
H^{(0)}_{\gamma = 0 , {\rm crit} }  = \sum_{ \alpha  = 1,2,3} \: \left\{ 
 - \frac{ J  }{2} \sum_{ j = 1}^{\ell - 1 } 
[ a_{j ,\alpha }^\dagger - a_{j , \alpha} ] [a_{j+1 , \alpha }^\dagger + a_{j+1 , \alpha } ]  
- \frac{J}{2} \sum_{ j = 1}^\ell [ a_{j , \alpha}^\dagger  + a_{j , \alpha } ] [a_{ j , \alpha }^\dagger - a_{j , \alpha } ] \right\}
\:\:\:\: .
\label{qic.1}
\eneq
\noindent
Upon defining the real-fermion vector operator $\vec{\Psi}_{j  }$ so that 

\begin{eqnarray}
\vec{\Psi}{j } &=& \left[ \begin{array}{c} a_{ j ,  1}^\dagger + a_{j , 1} \\ a_{j , 2}^\dagger + a_{j , 2} \\
a_{ j , 3}^\dagger + a_{j , 3} \end{array}\right] \;\;\; , \;\;  j \:{\rm odd} \nonumber \\
\vec{\Psi}_j &=& \left[ \begin{array}{c}- i ( a_{j ,1}^\dagger - a_{j , 1} ) \\ 
- i (a_{j , 2}^\dagger - a_{j , 2} ) \\ - i ( a_{j , 3}^\dagger - a_{j , 3} ) \end{array}
\right] \;\;\; , \;\; j \; {\rm even}
\:\:\:\: , 
\label{qic.2}
\end{eqnarray}
\noindent
with $j = 1 , \ldots , 2 \ell$, one sees that the definition in Eqs.(\ref{qic.2}) allows for rewriting $H_{\gamma = 0 , {\rm crit} } = 
H_{\gamma = 0 , {\rm crit}}^{(0)} + H_{\Delta , \gamma = 0}$ as 

\beq
H_{\gamma = 0 , {\rm crit}} = - \frac{iJ}{2} \: \sum_{j = 1}^{2 \ell - 1} \: \vec{\Psi}_j \cdot \vec{\Psi}_{j + 1} 
- 2 i J_\Delta ( \vec{\Psi}_1 \times \vec{\Psi}_1 ) \cdot \vec{\cal R}
\:\:\:\: . 
\label{qic.3}
\eneq

As outlined  in Ref.\cite{coleman}, the right-hand side of Eq.(\ref{qic.3}) corresponds to the 
lattice version of the real-fermion realization of a two-channel Kondo model. As the details of the renormalization 
group flow of the running coupling, of the nature of the strongly-coupled Kondo fixed point and of possible ways of 
experimentally probing the Kondo physics have been largely  addressed in the literature 
(see, for instance Refs.\cite{coleman,Giuliano_2013,gstt}), here we just stress once more 
how this remarkable  properties are determined by the emergence of the KF at the junction
and by their interaction  with the dynamical degrees of freedom of the junctions. At the same time, the topological nature of the spin 
operator $\vec{\cal R}$ is witnessed by the fact that  its components are nonlocal functionals of operators related to different leads 
of the junction, so that once one lead is disconnected, the operator itself ceases to exist as an effective quantum
spin-1/2 degree of freedom \cite{beri12}. 

For generic values of $\gamma$ and $H$, each single-chain fermionic Hamiltonian at the right-end
side of Eq.(\ref{xy.10}) may be readily diagonalized, yielding the dispersion relation  \cite{gstt}

\beq
\pm \epsilon_k = \pm \sqrt{ [ J ( 1 + \gamma ) \cos ( k ) - H ]^2 + J^2 ( 1- \gamma )^2 \sin^2 ( k ) }
\:\:\:\: . 
\label{bove.1}
\eneq
\noindent
For generic values of the parameters, the dispersion relation in Eq.(\ref{bove.1}) is fully gapped, 
with the gap $\Delta_w = \min \: \left\{ J ( 1 - \gamma ) \sqrt{ 1 - \frac{H^2}{\gamma^2 J^2} }  , 
| J ( 1 + \gamma ) - H | \right\}$. Moving along the "critical" line $\Delta_w = 0$, one may continuously 
interpolate between the Y-junction of QICs and the Y-junction of $XX$ spin chains 
studied in Ref.\cite{crampettoni}. 

\begin{figure}[h!]
 \centering
 \includegraphics[width=0.5\textwidth]{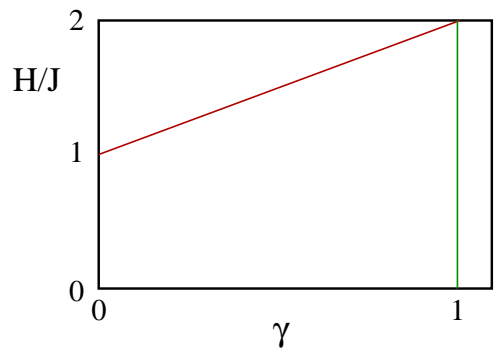}
 \caption{Phase diagram of the $XY$-chain in a magnetic field in 
 the $\gamma-(H/J)$ plane: the green line at $\gamma=1$ corresponds to the $XX$ line.
 The red line corresponds to the gap closure at  $H/J = 1+\gamma$ (see text). } 
\label{phad}
\end{figure}

In order to pertinently interpolate between the two Kondo systems 
discussed above, both characterized by a gapless dispersion relation of the 
lead fermions, one has to move along a line in parameter space on which $\Delta_w = 0$. 
This is explicitly illustrated in Fig.\ref{phad}.
The line continuosly connecting 
the two-channel Kondo system at $\gamma = 0$ with the four-channel one at $\gamma = 1$ is drawn in red
and while the vertical line with $\gamma=1$ is green.
 
Specifically, the ''critical'' line connecting the Ising limit $\gamma = 0$ with the four-channel line, spanned at 
constant $\gamma = 1$ by varying the ratio $ H / (2J)$ between -1 and 1, satisfies the equation $ H = J ( 1 + \gamma )$, 
which implies 

\beq
\pm \epsilon_{k , {\rm crit}} = \pm 2 J \left| \sin \left( \frac{k}{2} \right) \right| \: \sqrt{ ( 1 + \gamma )^2 \sin^2 
\left( \frac{k}{2} \right) + ( 1 - \gamma )^2 \cos^2 \left( \frac{k}{2} \right) } 
\:\:\:\: , 
\label{bove.2}
\eneq
\noindent
with the gap closing at $k = 0$. 

An important observation is that, though, for any $0 \leq \gamma < 1$, 
$H_\Delta$ corresponds to a four-channel, spin-1/2 Kondo Hamiltonian, the condition 
$J_2 / J_1 = \gamma < 1$ makes the coupling to be pairwisely inequivalent. This implies that 
$J_1$ will flow towards strong coupling before $J_2$. Once $J_1$ has flown to strong coupling, 
the flow of the other running coupling stops and the system behaves as a two-channel 
Kondo model.

\subsection{Thermodynamics of the topological Kondo model}
\label{appe}

The thermodynamics of the TKM with an arbitrary
number of wires $M$ is analyzed in Ref.\cite{Buccheri2015}, where it 
is provided the complete Bethe ansatz solution for $T \neq 0$. The results
for $T=0$ are given in Refs.\cite{Ogievetsky1987},\cite{DeVegaO2N}.
We refer to Ref.\cite{Buccheri2015} for details of the finite temperature  Bethe ansatz solution. 
Here, we limit ourselves to stress the most important results for the entropy and the specific heat.
  
The residual entropy at zero temperature shows that
the degrees of freedom introduced by the Majorana modes, contribute as  

\begin{equation}
\label{eq:DK-residualEntropy}
 S^{(0)}_J=\log \sqrt{\frac{M}{2}}
 \;,
\end{equation}
for even $M$ and
\begin{equation}\label{eq:BK-residualEntropy}
 S^{(0)}_J=\log \sqrt{M}
 \;,
\end{equation}
for odd $M$, in both cases
in agreement with the boundary conformal field theory results of
\cite{Altland2014}.

The signature of the non-Fermi liquid nature of the strongly coupled fixed point
is given by the next-to-leading term in the expansion of the
junction free energy. As a result, the Majorana contribution to the specific heat 
behaves at low temperatures as \cite{Buccheri2015}

\begin{equation}
\label{eq:specificHeat}
 C_J=-T\frac{\partial^2 F_J}{\partial T^2}\sim \left(\frac{T}{T_K}\right)^{\frac{2(M-2)}{M}}
 \;,
\end{equation}
where the (dimensionless) Kondo temperature $T_K$ depends on the coupling between legs as
\begin{equation}\label{eq:KondoTemperature}
 T_K \sim e^{-\frac{\pi}{\lambda(M-2)}}
 \;.
\end{equation}
A non-integer power is a strong and experimentally detectable
signature of the presence of a non-Fermi liquid fixed
point. In particular, it is related to the operator content of the
conformal field theory describing the fixed point at strong coupling,
as explained in Ref.\cite{Affleck1991641}.

\section{Josephson junction networks and junctions of spin chains}
\label{jjn}

In section \ref{sec:qsc} we have shown how real-fermion Majorana modes can emerge at 
a YSC. Spin chain models typically provide simplified, 
though effective, descriptions of strongly correlated,  many-body systems. 
A basic question to answer, when proposing a YSC as a system to probe emerging 
Majorana modes, is how to realize it in practice with controlled parameters. 
A possible answer to this question is provided by the mapping between 
arrays of Josephson junction chains   (JJC)s  with pertinently chosen parameters and 
quantum spin chains \cite{seb,gla,giuso_x}. In the following we remind how the mapping works for $XY$ model.

\subsection{The $XY$ model as an effective description of a Josephson junction chain }
\label{xyjj}

Starting from the pioneristic work by Bradley and Doniach \cite{seb}, it is by now 
well-estabilished how JJCs can be well-described in terms of either classical, 
or quantum, spin chains, in various regions of values of their parameters  \cite{seb,gla,giuso_x}.

The first relevant example is provided by a quantum $XX$-chain realized as a 
JJC   with finite charging energy $E_Q$. 

One considers a  chain of quantum Josephson junctions realized 
between superconducting grains, each one characterized by the value $\phi_i$ of 
the local superconducting order parameter.  The charge operator at grain-$i$ 
is canonically conjugated to $\phi_i$ and, in units of the Copper pair 
charge $e^* = 2e$, it is given by $\hat{Q}_i = - i \frac{\partial}{\partial \phi_i}$. 
Additionally, one  assumes that the charge at each grain is 
fixed by a gate voltage $V_g$. As a result, letting $E_Q , E_J $ and ${\cal N}$ respectively 
be equal to the charging energy at each grain, to the Josephson coupling between nearest-neighbor 
grains, and to the $V_g$-depending average charge at each grain, a uniform JJC with $\ell$ sites 
is described by the Hamiltonian \cite{gla,giuso_x}

\beq
H_{\rm JJC} = \frac{E_Q}{2} \: \sum_{ j = 1}^\ell \: \left[ - i \frac{\partial}{\partial \phi_j} - {\cal N} \right]^2 
- E_J \sum_{ j = 1}^{\ell - 1} \: \cos [ \phi_j - \phi_{j + 1} ] 
\:\:\:\: . 
\label{jj.1}
\eneq

To map $H_{\rm JJC}$ onto the model Hamiltonian for a quantum spin-1/2 $XX$-spin chain,  one should 
assume that the JJC parameters are set so that $E_Q / E_J \gg 1$. When ${\cal N}$ is integer, or near so, 
this corresponds to the so-called 
``charging'' regime of the JJC, in which, typically, the chain behaves as a (Mott) insulator, as 
the Coulomb blockade, due to the large charging energy, prevents charges from tunneling from grain 
to grain,  forbidding current transport across the chain \cite{seb}. 
In fact, at integer ${\cal N}$, the Coulomb blockade is determined by the condition  that the 
minimum energy state at each grain is nondegenerate, with total charge  $ \sim e^* {\cal N}$.
At variance, when ${\cal N} = n + \frac{1}{2}$, with integer $n$, two different charge 
eigenstates, with charge equal to $e^* n $ and to $e^* (n + 1)$, are degenerate, at each 
site of the chain. This means that one may recover the low-energy dynamics of $H_{\rm JJC}$ in 
Eq.(\ref{jj.1}) by only retaining those two states at each site. Accordingly, one naturally 
resorts to a quantum spin-1/2 re-formulation of the JJC dynamics, by defining, at each site $i$, the two states
$ | \uparrow \rangle_i \equiv | n+1 \rangle_i$, and $ | \downarrow \rangle_i = | n \rangle_i$, with 
$ | n + 1 \rangle_i $ and $ | n \rangle_i$ being the two charge eigenstates corresponding to 
charge $n+1$ and $n$ at site $i$. 

Defining the low-energy subspace as ${\cal F} = 
\oplus_{ \sigma_1 , \ldots , \sigma_\ell = \pm 1} \: \{ | \sigma_1 \rangle_1 \otimes \ldots \otimes | \sigma_\ell \rangle_\ell \}$, 
one defines the spin-1/2 operators $\vec{S}_j$ by projecting on ${\cal F}$ the operators at the right-hand side 
of Eq.(\ref{jj.1}). In particular, letting ${\cal P}_{\cal F}$ be the projector on ${\cal F}$, one obtains 
\cite{giuso_x}

\begin{eqnarray}
 S_j^z &\equiv& {\cal P}_{\cal F} \left[ - i \frac{\partial}{\partial \phi_j} - n - \frac{1}{2} \right] {\cal P}_{\cal F} \nonumber
 \\
 S_j^\pm &\equiv& {\cal P}_{\cal F} e^{ \pm i \phi_j } {\cal P}_{\cal F}
 \:\:\:\: . 
 \label{jj.2}
\end{eqnarray}
\noindent
In terms of the spin-1/2 operators in Eq.(\ref{jj.2}) one then gets 

\beq
H_{\rm JJC} = - \frac{E_J}{2} \: \sum_{ j =1}^{\ell - 1} \: \{ S_j^+ S_{ j+1}^- + S_{j+1}^+ S_j^- \} 
- h \sum_{ j = 1}^\ell S_j^z 
\:\:\:\: , 
\label{jj.3}
\eneq
\noindent
where, in Eq.(\ref{jj.3}), $h = E_Q \delta$, and $\delta$ corresponds to a possible small 
offset in ${\cal N}$ from the exact degeneracy point, that is, ${\cal N} = n + \frac{1}{2} + 
\delta$. 

One should observe that, though states with charge at a site $i$ different 
from either $n$, or $n+1$, are ruled out from ${\cal F}$ as ``high-energy'' states, at finite $E_Q$ they 
can still play a role as virtual states, entering the mapping leading to Eq.(\ref{jj.3}) by means 
of higher-order contributions in $E_J / E_Q$. For instance, to first-order in $E_J / E_Q$ an additional 
term at the right-hand side of Eq.(\ref{jj.3}) is generated, that is $\propto - \frac{E_J^2}{E_Q} 
\: \sum_{ j = 1}^{\ell - 1 } S_j^z S_{j+1}^z$ \cite{gla,giuso_x}. While such a term is of great interest in view of 
novel phase transitions it can trigger, its analysis goes beyond the scope of this work.  
The previous derivation shows how a JJC with appropriate parameters may be simulated by 
a   quantum $XX$-spin chain.

  At variance, engineering a Josephson junction  network realizing a QIC is much more challenging. 
In fact,  Cooper-pair tunneling between superconducting grains across each Josephson junction, 
naturally determines an $XX$ planar coupling, once one resorts to the spin-chain description of 
the JJC. In order to recover an Ising-like coupling in spin space it is required to use 
 a Josephson junction network, where the ``elementary unit'' is a rhombus, made out of a single, circular, 
four-junction chain \cite{Giuliano_2013}.

\subsection{Josephson junction network realization of Y-junctions of quantum spin chains}
\label{Yjos}

As a general observation, it is worth stressing again how, due to the optimal level of control 
reached on their fabrication and control parameters, Josephson junction networks provide an excellent arena to 
engineer reliable and largely tunable quantum devices \cite{havil}. 
In particular, it is by now known that highly coherent two-level quantum systems may 
emerge at pertinently engineered   Josephson junction rhombi chains  \cite{giusod_1,giusod_2}. 
More generally, it is well estabilished how a Josephson junction rhombi chain is able to induce  
charge $4e$ superconducting correlations \cite{rizzi,proto_1,proto_2}, in the bulk as well as at  a tunneling process 
across a quantum impurity  \cite{doucot,giusod_1,giusod_2}. Once one has set the mapping between 
a Josephson junction rhombi chain and a QIC, three Josephson junction rhombi chains   may be glued together into a
Y-junction, as shown in Ref.\cite{Giuliano_2013},
thus providing the JJC realization of Tsvelik's YSC.

Given the mapping between pertinent Josephson junction networks and quantum spin chains 
($XX$-spin chains or QICs), one expects to translate all the observations concerning 
the emergence of Majorana modes at a YSC to Y-junctions realized with appropriate 
Josephson junction networks. To illustrate how the task can be achieved, one begins 
with a Y-junction of JJCs such as the ones described in Eq.(\ref{jj.1}).
One considers three chains, described by the Hamiltonian $H_{\rm JJC}^{(3)}$, given by 

\beq
H_{\rm JJC}^{(3)}  = \sum_{\alpha = 1,2,3} \: \left\{ \frac{E_Q}{2} \: \sum_{ j = 1}^\ell \: \left[ - i \frac{\partial}{\partial \phi_{ j , \alpha} } - {\cal N} \right]^2 
- E_J \sum_{ j = 1}^{\ell - 1} \: \cos [ \phi_{j , \alpha}  - \phi_{j + 1 , \alpha} ] \right\}  
\:\:\:\: ,
\label{jjy.1}
\eneq
\noindent
with $\phi_{j , a}$ being the phase of the superconducting grain $j$ of chain $\alpha$ and $- i \frac{\partial}{\partial \phi_{j, \alpha} } $
being the corresponding charge operator. The three chains are connected at one of their endpoints, say $j=1$, 
to the other two chains by means of a Josephson coupling term $H_J$, given by \cite{crampettoni}

\beq
H_J = - J' \: \sum_{\alpha = 1,2,3} \: \{ e^{ i \phi_{1 , \alpha} } e^{ - i \phi_{1,\alpha+1} } +  e^{  i \phi_{1,\alpha +1} } e^{ -  i \phi_{1 , \alpha} }  \} 
\:\:\:\: . 
\label{jjy.2}
\eneq

Assuming that the parameters are all the same in the ``bulk'' of each chain, one may undergo through the same 
projection onto the joint low-energy subspace of the three chains, thus resorting to an effective, spin-1/2 
quantum spin chain description of the network. As a result, setting ${\cal N} = \bar{N} + \frac{1}{2} + h$, with 
$| h | \ll 1$,  ${\cal F}_a\alpha= \oplus_{ \sigma_{1 , \alpha }  , \ldots , \sigma_{\ell , \alpha } = \pm 1} \: 
\{ | \sigma_{1 , \alpha } \rangle_{1,\alpha} \otimes \ldots \otimes | \sigma_{\ell , \alpha } \rangle_{\ell , \alpha } $ and  
${\cal F} = \prod_{\alpha = 1,2,3} \: {\cal F}_\alpha $, one defines $S_{j,\alpha}^z = {\cal P}_{\cal F} 
\left[ - i \frac{\partial}{\partial \phi_{j , \alpha } } - \bar{N} - \frac{1}{2} \right] {\cal P}_{\cal F}$,
$S_{j , \alpha}^\pm  = {\cal P}_{\cal F} e^{ \pm i \phi_{j , \alpha }} {\cal P}_{\cal F}$. Then one obtains 
that the whole junction (the set of the three JJCs plus the coupling at the endpoints of the chains) is 
described by the Hamiltonian $H_{\rm junction}$, given by 

\begin{eqnarray}
&& H_{\rm junction} = {\cal P}_{\cal F} [ H_{\rm JJC}^{(3)}  + H_J ] {\cal P}_{\cal F} =  \\
&& \sum_{\alpha = 1,2,3} \: \left\{ - \frac{E_J}{2} \: \sum_{ j = 1}^{\ell - 1 } [ S_{ j , \alpha }^+ S_{j + 1 , \alpha }^- + 
S_{j + 1 , \alpha }^+ S_{j , \alpha }^- ] - h \sum_{j = 1}^\ell S_{j , \alpha }^z \right\} 
- 2 J_\Delta  \sum_{ \alpha = 1,2,3} S_{1,\alpha}^x S_{1 , \alpha+1}^x \nonumber 
\:\:\:\: ,
\label{jjy.3}
\end{eqnarray}
\noindent
with $ J_\Delta / E_J < 1$, that is, the Hamiltonian of a YSC in the isotropic case $\gamma = 1$. 

One may construct a YSC of quantum Ising chains, as illustrated in \cite{Giuliano_2013}, 
by considering three equal rhombi chains described by the Hamiltonian $H_{\rm m}^{(3)}$ given by 

\begin{eqnarray}
H_{\rm m}^{(3)}  &=& \sum_{\alpha=1,2,3} \Biggl\{  - J \sum_{ p = 1}^\ell  \sum_{ r =1}^4 \{ e^{-\frac{i}{4} \varphi} \sigma_{r , p , \alpha }^+ \sigma_{
r + 1 , p , \alpha }^- + {\rm h.c.} \} \nonumber \\
& -& 
h\sum_{p=1}^\ell   \sum_{ r = 1}^4 \sigma_{r , p , \alpha }^z - {\cal T}   \sum_{ p = 1}^{\ell - 1}
\{ \sigma_{3 , p , \alpha }^+ \sigma_{1 , p+1 , \alpha }^- + {\rm h.c.} \} \Biggr\} 
\:\:\:\: .
\label{jjy.4}
\end{eqnarray}
\noindent
To effectively make a Y-junction, one assumes the existence of  Josephson couplings with 
strength ${\cal J}$ between  sites number 2 of the endpoint-rhombus of
each chain, described by the Hamiltonian $H_{\rm MB;J}$ given by

\beq
H_{\rm MB;J} = - {\cal J} \{ \sigma_{1,1,2}^+ \sigma_{2,1,2}^- + \sigma_{2,1,2}^+ \sigma_{3,1,2}^- + \sigma_{3,1,2}^+ \sigma_{1,1,2}^- + {\rm h.c.} \}
\:\:\:\: .
\label{jjy.5}
\eneq
\noindent
One then projects with ${\cal P}_{\cal G}$ the total Hamiltonian $H_{\rm m}^{(3)} + H_{\rm MB;J}$.
As a result, one gets the Hamiltonian $H_Y$, given by 

\beq
H_{Y} = \sum_{ \lambda = 1,2,3} \{
J_x \sum_{ p = 1}^{\ell - 1}  S_{p , \lambda}^x S^x_{p+1 , \lambda} - 2  \sum_{p = 1}^\ell H_p  S_{p , \lambda}^z
\}+ H_{B}
\:\:\:\: ,
\label{jjy.6}
\eneq
\noindent
with  

\beq
H_{B} = J_K  \{ S_{1,1}^x S_{2,1}^x + S_{2,1}^x S_{3,1}^x + S_{3,1}^x S_{1,1}^x\} + \delta H_B
\:\:\:\: ,
\label{jjy.7}
\eneq
\noindent
and $J_K = \frac{{\cal J}^2}{J}$. 
$\delta H_B = - \frac{\sqrt{2}}{16} \frac{{\cal J}^2}{J}  \{ S_{1,1}^z + S_{2,1}^z + S_{3,1}^z \}$ is a
boundary magnetic field term which does not affect the behavior of the system. Apart for this term
(and for a change in sign of $J_x$), the Hamiltonian in 
Eqs.(\ref{jjy.6})-(\ref{jjy.7}) is exactly the one describing the YSC of QICs introduced before. 

\section{Ultracold atoms and junctions of spin chains}
\label{sec:hol}

In order to implement a $Y$-junction
with cold atomic systems \cite{Cassettari}, one has to recall that:

\begin{itemize}

\item[{\em 1}] a Tonks-Girardeau   (TG) gas of one-dimensional bosons on a $Y$-junction is a physical
  realization of the $XX$ model on the $Y$-junction itself -- and therefore can be mapped
  via the JW transformation of Section \ref{sec:qsc} in the TKM Hamiltonian;

\item[{\em 2}] TG gases of ultracold bosons have been implemented in several experiments  \cite{Cazalilla11};

\item[{\em 3}] Stable $Y$-configurations may be created by holographic techniques, as experimentally shown in \cite{Cassettari}.
  
  \end{itemize}

For our purpose here, one should focus only on  point {\em 1)} and 
  show  that (\ref{eq:TopologicalKondoHamiltonian}) may  be 
obtained from a model Hamiltonian for interacting bosons  in the TG limit, confined to $M$ one-dimensional waveguides 
arranged in a $Y$-junction. 

In each waveguide $\alpha=1,\cdots,M$ the Lieb-Lininger Hamiltonian describing interacting
bosons in one-dimensional guides of length $\mathcal{L}$
reads \cite{Lieb63,Yang69,korepin1997quantum}:
\begin{equation}
H^{(\alpha)}=\int_{0}^{\mathcal{L}}dx\left[\frac{\hbar^{2}}{2m}\partial_{x}
\Psi_\alpha^{\dagger}(x)\partial_{x}\Psi_\alpha(x)+\frac{c}{2}
\Psi_\alpha^{\dagger}(x)\Psi_\alpha^{\dagger}(x)\Psi_\alpha(x)\Psi_\alpha(x)\right]\;.
\label{eq:LiebLininger}
\end{equation}
The parameter $m$ is the mass of the bosons and $c>0$ is the repulsion strength,
as determined by the $s$-wave scattering length \cite{olshanii}.
The bosonic fields $\Psi_\alpha$ satisfy canonical commutation relations 
$\left[\Psi_\alpha(x),\Psi_\alpha^{\dagger}(y)\right]=\delta(x-y)$. 

The coupling of the Lieb-Lininger Hamiltonian, denoted by $\gamma$, is proportional 
to $c/n$ where $n\equiv\mathcal{N}/\mathcal{L}$ is the density
of bosons and $\mathcal{N}$ is the number of bosons per waveguide. More specifically one has  $\gamma=mc/\hbar^2 n$. 
The limit of vanishing $\gamma$ corresponds to an ideal one-dimensional Bose 
gas, while the limit of infinite $\gamma$ corresponds to 
the TG gas \cite{Tonks36,Girardeau60}, which generally has the 
expectation values and thermodynamic quantities of a one-dimensional ideal Fermi gas 
\cite{korepin1997quantum,Yurovsky08,Bouchoule09,Cazalilla11}.
The experimental realization of the TG gas with cold atoms 
\cite{Paredes2009,Kinoshita2004} triggered intense 
activity in the last decade, reviewed in \cite{Yurovsky08,Bouchoule09,Cazalilla11}. 

One considers $M$ copies of this one-dimensional Bose gas and 
joins them together by the ends of the
segments, in such a way that the bosons can tunnel from one waveguide 
to the others. The bosonic fields in different  
legs commute:
$$\left[\Psi_\alpha(x),\Psi_\beta^{\dagger}(y)\right]=
\delta_{\alpha,\beta} \,  \delta(x-y),$$ and the total Hamiltonian has the form 
$H=\sum_{\alpha=1}^{M} H^{(\alpha)}+H_J$ where the junction term 
$H_J$ describes the tunneling process among legs.

As a tool for performing computations, as well as to give a precise
meaning to the tunneling processes at the edges of the legs, 
in each leg we discretize space into a lattice of 
$L$ sites with lattice spacing $a$ (where $La=\mathcal{L}$ and the total 
number of sites $N_S$ of the star lattice is $N_S \equiv LM$). 
This discretization 
can be physically realized by superimposing optical lattices on the legs
\cite{LewensteinBook}. One can then perform a tight-binding approximation 
\cite{Jaksch98,Trombettoni01} and write the
bosonic fields as $\Psi_\alpha(x)=\sum_{\alpha, j} 
w_{\alpha, j}(x) \, b_{\alpha, j}$ where 
$b_{\alpha, j}$ is the operator destroying a particle in the site $j=1,\cdots,L$ 
of the leg $\alpha$ and $w_{\alpha, j}(x)$ 
is the appropriate Wannier wavefunction 
localized in the same site.

The resulting lattice Bose-Hubbard 
Hamiltonian on each leg then reads \cite{Jaksch98,Jaksch2005}
\begin{equation}
H_{U}^{(\alpha)} = -t\sum_{j=1}^{L-1}\left(b_{\alpha, j}^{\dagger}b_{\alpha, j+1}+
b_{\alpha, j+1}^{\dagger}b_{\alpha, j}\right)+
\frac{U}{2}\sum_{j=1}^{L}b_{\alpha, j}^{\dagger}b_{\alpha, j}^{\dagger}
b_{\alpha, j}b_{\alpha, j}\label{eq:BosonSingleChainHamiltonian}
\end{equation}
where the interaction coefficient is $U=c\int\left|w_{\alpha}(x)\right|^{4}dx$
($\alpha=1,\cdots,L$), the hopping coefficient 
is $t=-\int w_{\alpha,j} \hat{T} 
w_{\alpha,j+1} \,dx$ with $\alpha=1,\cdots,L-1$, and 
$\hat{T}=-(\hbar^2/2m)\partial^2/\partial  x^2$ is the kinetic energy operator.

The total lattice Hamiltonian for a $Y$ junction  of atomic condensates  is obtained by taking 3 copies of 
the system, 
connected to one another by a hopping term. The total Hamiltonian is then written
as:
\begin{equation}\label{totalHam}
H_U=\sum_{\alpha=1}^{3}H_{U}^{(\alpha)}+H_J,
\end{equation}
where the junction term has the form 
\begin{equation}
H_J=-\lambda\sum_{1\le\alpha<\beta\le 3}\left(b_{\alpha,1}^{\dagger}b_{\beta,1}+b_{\beta,1}^{\dagger}b_{\alpha,1}\right)\label{eq:BosonHamiltonian}
\end{equation}
with $\lambda$ being the hopping between the first site of a leg 
and the first sites of the others. 
Typically one has $\lambda>0$, which corresponds to an {\it antiferromagnetic} Kondo model,
as shown in the following.
Nevertheless, we observe that the sign of $t$ and
$\lambda$ could be changed by shaking the trap \cite{Eckardt}.

The total number of bosons 
in the system, $N=\mathcal{N}\mathcal{L}$, is a conserved quantity
in the lattice model and can be tuned in experiments. 
In the canonical ensemble $N=\sum_{\alpha,j} \left\langle b_{\alpha, j}^{\dagger} b_{\alpha, j} \right\rangle$. 
The phase diagram of the bulk Hamiltonian 
(\ref{eq:BosonSingleChainHamiltonian}) in each leg 
undergoes quantum phase transitions between superfluid and Mott insulating 
phases \cite{Fisher1989}: notice that in the canonical ensemble the system 
is superfluid as soon as the filling $N/N_S$ is not integer.

One is interested in the limit $U\to\infty$, so that 
after the continuous limit is taken back again, the TG gas 
is retrieved in the bulk. It is well known that this limit brings
substantial simplifications 
in the computation: it was shown in \cite{Friedberg199352}
that, on each leg, the spectrum and the scattering matrix are equivalent 
to a system of spins in the $s=1/2$
representation. As customary, one may  map the hard-core bosons to  $1/2$ spins. 
The Hamiltonian (\ref{totalHam})   written in spin variables is given by 
\begin{eqnarray}
H_{\infty}^{(\alpha)} & = & -t\sum_{j=1}^{N-1}\left(S_{j , \alpha}^{+}\sigma_{j+1 ,\alpha}^{-}+S_{j+1,\alpha}^{+}\sigma_{j,\alpha}^{-}\right)\label{eq:SpinHamiltonian_a}\\
H_{J} & = & -\lambda\sum_{\alpha<\beta}^{3}\left(S_{1,\alpha}^{+}\sigma_{1,\beta}^{-}+\sigma_{1,\beta}^{+}\sigma_{1,\alpha}^{-}\right)\label{eq:SpinHamiltonian} 
\end{eqnarray}
which coincides with a junction of $XX$-type spin chains \cite{crampettoni}. Now one can proceed exactly as in Section \ref{sec:qsc}, finding
an Hamiltonian of the form (\ref{xy.delta}) via the correct  identification of the parameters.

From the Hamiltonian 
(\ref{eq:SpinHamiltonian_a})-(\ref{eq:SpinHamiltonian}), one obtains
\begin{eqnarray}
H & = & -t\sum_{j=1}^{N-1}\sum_{\alpha=1}^{3}\left(c_{\alpha,j}^{\dagger}c_{\alpha,j+1}+c_{\alpha,j+1}^{\dagger}c_{\alpha,j}\right)+H_J\label{eq:TOTFER} \\
H_J & = & -\lambda\sum_{1\le\alpha<\beta\le 3}\gamma_{\alpha}\gamma_{\beta}\left(c_{\alpha,1}^{\dagger}c_{\beta,1}+c_{\alpha,1}c_{\beta,1}^{\dagger}\right)\label{eq:LatticeFermionHamiltonianM}
\end{eqnarray}
In conclusion, one has  mapped the Hamiltonian (\ref{eq:SpinHamiltonian}),
acting on $N_S$ spin variables, onto another one, defined in terms
of $N_S$ spinless fermionic degrees of freedom plus one Klein factor per leg. 
In other words, the hard-core boson Hamiltonian (\ref{totalHam}) in the limit $U\to\infty$
is mapped onto
the fermionic Hamiltonian (\ref{eq:TOTFER}), given by the sum of 
non-interacting wires and the highly nontrivial junction term $H_{J}$. 
The Fermi energy of the non-interacting fermions in (each of) 
the external wires is denoted by $E_F$.

It should be remarked that the only mechanism in which the topological protection of the degree of freedom
encoded in the Majorana modes can be spoiled is the loss of bosonic atoms by the trap. With our parameters,
we found this probability  negligible on the time scales needed for the experiment, 
since the energy barrier for the atom loss is $\sim 500$~nK, much larger than the typical Fermi energy of the 
TG gases.

\section{Concluding remarks}
\label{sec:conc}

We argued that a pertinent modification of a remarkable result obtained in the seventies by Roman Jackiw and collaborators 
\cite{JackiwRebbi,JackiwRossi} opened the avenue to consistently deal with non-local effects in   field theories of relevant condensed matter systems. 

In this paper we dealt with these non-local effects in spin models relevant for the study of the multi-channel Kondo effect.
One advantage of these Kondo models is that the perfect symmetry of the couplings of the various channels with the spin impurity is guaranteed 
by a topological degeneracy and thus robust against decoherence.
The topological Kondo effect stems from the interaction of localized Majorana modes with external 1D channels arranged in a pertinent graph geometry.

 We considered the simple case of star-like geometries on which $XX$ and $XY$ spin models were defined. We showed how, from these models,  
 the four- and two-channel Kondo models emerge.
 
We finally showed  how these topological Kondo Models may be realized in Josephson Junction arrays and networks of Tonks-Girardeau gases.

\section*{Acknowledgements:}

 We all benefited from  discussions   with Professor Roman Jackiw and from his
lectures on Majorana fermions and charge fractionalization. 
We are very happy to contribute to His festschrift volume.

I (P.S.) enjoyed Roman'€™s friendship since 1985 when I, firstly, joined His research group at the Centre for theoretical physics of MIT. 
The possibility to visit MIT as a postdoctoral fellow in Roman's group was offered to me by Professor Sergio Fubini who, after
the many discussions we had at CERN, assisted me in getting the financial support for my first visit at MIT. There, it started a very
memorable time of my scientific and personal life! Indeed, I had the opportunity to enjoy  the many teachings from Roman on a variety 
of topics ranging from Chern-Simons field theories to field theories of condensed matter systems. Our frequent  discussions shaped my scientific 
attitude through the years and drove my interest towards the study of the role of topology in relativistic field theories and condensed 
matter systems. The friendship with Roman gave to me the unique chance to appreciate also His sense of humor,   His loyalty as a friend,
and His patience as a teacher; 
in addition, He had a unique ability in choosing always the very good restaurants where, from time to time, we celebrated our encounters. I wish 
Him a very happy birthday and many serene and interesting  years to come.

It is also a pleasure to thank I. Affleck, H. Babujian, A. Bayat, 
S. Bose, F. Buccheri, M. Burrello, G. Campagnano, N. Cramp\'e,
D. Cassettari, R. Egger, R. Graham, V. Korepin, L. Lepori, D. Rossini,  and A. Tagliacozzo
for many enlightening discussions during our collaboration
on some of the topics presented in this paper.
 \bibliographystyle{ws-rv-van}
 \bibliography{Y.bib}
 \end{document}